\documentclass[11pt]{article}
\usepackage{graphicx}
\usepackage{amsfonts}
\usepackage{theorem}
\newtheorem{Lem}{Lemma}[section]

\newtheorem{The}[Lem]{Theorem}
\newtheorem{Prop}[Lem]{Proposition}
\newtheorem{Cor}[Lem]{Corollary}

\newtheorem{Rem}[Lem]{Remark}

\newcommand{\qed}{\hbox{\rule{6pt}{6pt}}}

\begin{document}
\title{Characterizations of generalized entropy functions by functional equations}
\author{Shigeru Furuichi$^1$\footnote{E-mail:furuichi@chs.nihon-u.ac.jp}\\
$^1${\small Department of Computer Science and System Analysis,}\\
{\small College of Humanities and Sciences, Nihon University,}\\
{\small 3-25-40, Sakurajyousui, Setagaya-ku, Tokyo, 156-8550, Japan}}
\date{}
\maketitle
{\bf Abstract.} We shall show that a two-parameter extended entropy function is characterized by a functional equation. 
As a corollary of this result, we obtain that the Tsallis entropy function is characterized by a functional equation,
which is a different form used in \cite{ST} i.e., in Proposition \ref{prop01} in the present paper. 
We also give an interpretation of the functional equation giving the
Tsallis entropy function, in the relation with two non-additive properties.
\vspace{3mm}

\vspace{3mm}


\vspace{3mm}


\section{Introduction}
Recently, generalized entropies have been studied from the mathematical point of view.
The typical generalizations of Shannon entropy \cite{Sha} are R\'enyi entropy \cite{Ren} and Tsallis entropy \cite{Tsa}.
The R\'enyi entropy  and the Tsallis entropy are defined by
$$
R_q(X) = \frac{1}{1-q} \log \sum_{j=1}^n  x_j^q, \quad (q\neq 1,q > 0)
$$
and 
$$
S_q(X)=  \sum_{j=1}^n \frac{x_j-x_j^q}{q-1},\quad (q\neq 1, q>0)
$$
for a given information source $X=\left\{x_1,\cdots,x_n\right\}$ with the probability $p_j\equiv Pr(X=x_j)$.
Both entropies recover Shannon entropy:
$$
S_1(X) \equiv -\sum_{j=1}^n p_j \log p_j
$$
in the limit $q \to 1$.
The uniqueness theorem for the Tsallis entropy was firstly given in \cite{Suy} and improved in \cite{Furu1}.

Throughout this paper, a parametric extended entropy such as R\'enyi entropy and Tsallis entropy and so on, is called by a generalized entropy.

We note that the R\'enyi entropy  has the additivity:
$$
R_q(X\times Y) =R_q(X) +R_q(Y)
$$
but the Tsallis entropy has the non-additivity:
\begin{equation}  \label{non-additive01}
S_q(X\times Y) =S_q(X) +S_q(Y) + (1-q)S_q(X)S_q(Y),
\end{equation}
where $X\times Y$ means that $X$ and $Y$ are independent random variables.
Therefore we have a definitive difference for these entropies, although we have the simple relation
between them:
$$
R_q(X) = \frac{1}{1-q} \log \left\{  1+(1-q)S_q(X) \right\},  \quad (q\neq 1).
$$ 

The Tsallis entropy is rewritten by 
\begin{equation}  \label{Tsa_re1}
S_q(X) = - \sum_{j=1}^n p_j^q \ln_q p_j 
\end{equation}
where the $q$-logarithmic function is defined by
$$
\ln_q x \equiv \frac{x^{1-q}-1}{1-q},\quad (q\neq 1),
$$
which uniformly converges to $\log x$ in the limit $q\to 1$.

Since Shannon entropy can be regarded as the expectation value for  each value $-\log p_j$,
we may consider that  the Tsallis entropy can be regarded as the $q$-expectation value for  each value $-\ln_q p_j$,
as an anology to Shannon entropy.
Where $q$-expectation  value $E_q$ is defined by
$$
E_q(X)\equiv \sum_{j=1}^n p_j^q x_j.
$$
However, the $q$-expectation  value $E_q$ lacks the fundamental property such as $E(1)=1$,
so that it was considered to be inadequate to adopt as a generalized definition of the usual expectation value.
Then the noremalized $q$-expectation value was introduced:
$$
E_q^{(nor)}(X) \equiv \frac{\sum_{j=1}^n p_j^q x_j}{\sum_{i=1}^np_i^q}
$$
 and by using this, the normalized Tsallis entropy was defined by
$$
S_q^{(nor)}(X) \equiv \frac{S_q(X)}{\sum_{j=1}^np_j^q} = - \frac{\sum_{j=1}^n p_j^q \ln_q p_j}{\sum_{i=1}^np_i^q},\quad (q\neq 1).
$$
We easily find that we have the following non-additivity relation for the normalized
Tsallis entropy:
\begin{equation}  \label{non-additive02}
S_q^{(nor)}(X\times Y) =S_q^{(nor)}(X)+S_q^{(nor)}(Y) +(q-1)S_q^{(nor)}(X)S_q^{(nor)}(Y).
\end{equation} 
As for the details on the mathematical properties of the normalized Tsallis entropy, see \cite{Suy0} for example.
The difference between two non-additivities Eq.(\ref{non-additive01}) and Eq.(\ref{non-additive02}) is the 
signature of the coefficient $1-q$ in the third term of the right hand sides.

We note that the Tsallis entropy is also rewritten by
\begin{equation}  \label{Tsa_re2}
S_q(X) =  \sum_{j=1}^n p_j \ln_q \frac{1}{p_j} 
\end{equation}
so that we may regard it as the expectation value such as $S_q(X)= E_1\left[\ln_{q}\frac{1}{p_j}\right]$, where $E_1$ means the usual expectation value.
However, if we adopt this formulation in the definition of the Tsallis conditional entropy, we do not have an important property such as a chain rule.
(See \cite{Furu2} for details.) Therefore we often adopt the formulation using the $q$-expectation value.

As a further generalization, for two positive numbers $\alpha$ and $\beta$,
a two-parameter extended entropy:
$$
S_{\alpha,\beta}(X) \equiv \sum_{j=1}^n \frac{x_j^{\alpha}-x_j^{\beta}}{\beta -\alpha}, \quad (\alpha \neq \beta)
$$
has been studied in many literatures \cite{STa,Mit,BR,KLS1,KLS2,WS,Furuichi}.

In the paper \cite{ST}, a characterization of the Tsallis entropy function was proven by using the functional equation.  
In this paper, we shall show that the two-parameter extended entropy function is characterized by the functional equation.


\section{A review of the characterization of Tsallis entropy function by the functional equation}
The following proposition was originally given in \cite{ST} by the simple and elegant proof.
Here we give the alternative proof along to the proof given in \cite{Horibe}.

\begin{Prop} \label{prop01}  {\bf (\cite{ST})}
If the differentiable nonnegative function $f_q$ with positive parameter $q\in \mathbb{R}$ satisfies the following functional equation:
\begin{equation} \label{func_eq01}
f_q(xy)+f_q((1-x)y)-f_q(y)=\left(f_q(x) +f_q(1-x)\right)y^q,\qquad (0< x < 1,\quad 0< y \leq 1)
\end{equation}
then the function $f_q$ is uniquely given by
$$
f_q(x) =-c_q x^q \ln_q x,
$$
where $c_q$ is a nonnegative constant depending only on the parameter $q$.
\end{Prop}

{\it Proof}:

If we put $y=1$ in Eq.(\ref{func_eq01}), then we have $f_q(1)=0$. From here we assume $y \neq 1$.
We also put $g_q(t)\equiv \frac{f_q(t)}{t}$. Then we have
\begin{equation} \label{eq01}
xg_q(xy)+(1-x)g_q((1-x)y)-g_q(y)=\left(xg_q(x)+(1-x)g_q(1-x) \right)y^{q-1}
\end{equation}
Putting $x=\frac{1}{2}$ in (\ref{eq01}), we have
$$
g_q\left(\frac{y}{2}\right)=g_q\left(\frac{1}{2}\right)y^{q-1}+g_q(y).
$$
Substituting $\frac{y}{2}$ into $y$, we have
$$
g_q\left(\frac{y}{2^2}\right)=g_q\left(\frac{1}{2}\right)\left(y^{q-1}+\left(\frac{y}{2}\right)^{q-1}\right) +g_q(y).
$$
By repeating similar substitutions, we have
\begin{eqnarray*}
g_q\left(\frac{y}{2^N}\right)&=&  g_q\left(\frac{1}{2}\right)    y^{q-1}   \left(1   +  \left(\frac{1}{2}\right)^{q-1}  +\left(\frac{1}{2}\right)^{2(q-1)} +\cdots +\left(\frac{1}{2}\right)^{(N-1)(q-1)}  \right) +g_q(y)\\
&=&  g_q\left(\frac{1}{2}\right)    y^{q-1}   \left(\frac{2^{N(1-q)-1}}{2^{1-q}-1}\right) +g_q(y).
\end{eqnarray*}
Then we have
\begin{equation} \label{eq02}
\lim_{N\to \infty} \frac{g_q\left(\frac{y}{2^N}\right)}{2^N}=0
\end{equation}
due to $q>0$.
Differentiating  Eq.(\ref{eq01}) by $y$, we have
$$
x^2g_q(xy)+(1-x)^2g_q((1-x)y)-g_q'(y)=(q-1)\left(xg_q(x)+(1-x)g_q(1-x)\right)y^{q-2}
$$
Putting $y =1$ in the above equation, we have
\begin{equation} \label{eq04}
x^2g'_q(x)+(1-x)^2g'_q(1-x)+(1-q)\left(xg_q(x)+(1-x)g_q(1-x)\right)=-c_q,
\end{equation}
where $c_q=-g_q'(1)$.

By integrating the equation (\ref{eq01}) from $2^{-N}$ to $1$ with respect to $y$ and performing the conversion of the variables,
we have
\begin{equation} \label{eq05}
\int_{2^{-N}x}^x g_q(t)dt + \int_{2^{-N}(1-x)}^{1-x}g_q(t)dt -\int_{2^{-N}}^1g_q(t)dt
=\left( xg_q(x)+(1-x)g_q(1-x) \right)\frac{1-2^{-qN}}{q}.
\end{equation}
By differentiating the above equation  with respect to $x$, we have

\begin{eqnarray*}
&&g_q(x)-2^{-N}g_q(2^{-N}x)-g_q(1-x)+2^{-N}g_q(2^{-N}(1-x))\\
&&=\frac{1-2^{-qN}}{q} \left( g_q(x)+xg_q'(x) -g_q(1-x)-(1-x)g'_q(1-x)\right).
\end{eqnarray*}
Taking the limit $N \to \infty$ in the above,
we have 
\begin{equation} \label{eq06} 
(1-x)g'_q(x)+(1-q)g_q(1-x)=xg'_q(x)+(1-q)g_q(x)
\end{equation}
thanks to (\ref{eq02}).
From the equations (\ref{eq04}) and (\ref{eq06}), we have the following differential equation:
$$
xg'_q(x)+(1-q)g_q(x)=-c_q.
$$
This differential equation has the following general solution:
$$
g_q(x)=-\frac{c_q}{1-q}+d_qx^{q-1},
$$
where $d_q$ is a integral constant depending on $q$.
From $g_q(1)=0$, we have $d_q=\frac{c_q}{1-q}$.
Thus we have
$$
g_q(x)=c_q\frac{x^{q-1}-1}{1-q}. 
$$
Finally we have 
$$
f_q(x)=c_q\frac{x^q-x}{1-q}=-c_qx^q\ln_qx.
$$
From $f_q(x) \geq 0$, we have $c_q \geq 0$.

\hfill \qed

If we take the limit as $q \to 1$ in Theorem \ref{prop01}, we have the following corollary.

\begin{Cor} \label{cor01}   {\bf (\cite{Horibe})}
If the differentiable nonnegative function $f$ satisfies the following functional equation:
\begin{equation} \label{func_eq11}
f(xy)+f((1-x)y)-f(y)=\left(f(x) +f(1-x)\right)y,\qquad (0< x < 1,\quad 0< y \leq 1)
\end{equation}
then the function $f$ is uniquely given by
$$
f(x) =-c x \log x,
$$
where $c$ is a nonnegative constant.
\end{Cor}


\section{Main results}
In this section, we give a characterization of a two-parameter extended entropy function by the functional equation.
Before we give our main theorem, we review the following result given by Pl.Kannappan \cite{Kan1,Kan2}.

\begin{Prop} {\bf (\cite{Kan1,Kan2})}
Let two probability distributions $(p_1,\cdots,p_n)$ and $(q_1,\cdots,q_m)$. 
If the measureable function $f:(0,1) \to \mathbb{R}$ satisfies 
\begin{equation}  \label{con_Kan}
\sum_{i=1}^n\sum_{j=1}^m f(p_iq_j) =  \sum_{i=1}^n p_i^{\alpha} \sum_{j=1}^m f(q_j) +  \sum_{j=1}^m q_j^{\beta} \sum_{i=1}^n f(p_i), 
\end{equation}
for all $(p_1,\cdots,p_n)$ and $(q_1,\cdots,q_m)$,
then  the function $f$ is given by
\[
f\left( p \right) = \left\{ \begin{array}{l}
 c\left( {p^\alpha   - p^\beta  } \right),\,\,\,\,\,\,\,\,\,\,\,\,\,\,\,\,\,\,\,\,\,\,\,\,\,\,\,\,\,\,\,\,\,\,\,\,\,\,\,\,\,\,\,\,\,\,\,\,\,\,\,\,\,\,\,\,\,\,\,\,\,\,\,\,\,\,\alpha  \ne \beta , \\ 
 cp^\alpha  \log p,\,\,\,\,\,\,\,\,\,\,\,\,\,\,\,\,\,\,\,\,\,\,\,\,\,\,\,\,\,\,\,\,\,\,\,\,\,\,\,\,\,\,\,\,\,\,\,\,\,\,\,\,\,\,\,\,\,\,\,\,\,\,\,\,\,\,\,\,\,\,\,\,\,\,\alpha  = \beta , \\ 
 cp\log p + b\left( {mn - m - n} \right)p + b,\,\,\,\,\,\,\,\,\,\,\,\,\,\,\,\,\alpha  = \beta  = 1. \\ 
 \end{array} \right.
\]
where $c$ and $b$ are arbitrary constants. 
\end{Prop}

In the following theorem, we adopt a simpler condition than Eq.(\ref{con_Kan}).

\begin{The} \label{the02}
If the differentiable nonnegative function $f_{\alpha,\beta}$ with two positive parameters $\alpha,\beta \in \mathbb{R}$ satisfies the following functional equation:
\begin{equation} \label{func_eq02}
f_{\alpha,\beta}(xy)  =  x^{\alpha} f_{\alpha,\beta}(y)+  y^{\beta} f_{\alpha,\beta}(x) , \qquad (0< x,y \leq 1)
\end{equation}
then the function $f_{\alpha,\beta}$ is uniquely given by
$$
f_{\alpha,\beta}(x) =c_{\alpha,\beta}  \frac{x^{\beta}-x^{\alpha}}{\alpha - \beta},\quad (\alpha\neq \beta)
$$
and
$$
f_{\alpha}(x) = -c_{\alpha} x^{\alpha}\log x, \quad (\alpha=\beta)
$$
where $c_{\alpha,\beta}$ and $c_{\alpha}$ are  nonnegative constants depending only on the parameters $\alpha$ (and $\beta$).
\end{The}
{\it Proof}:

If we put $y=1$, then we have $f_{\alpha,\beta}(1)=0$ due to $x>0$. 
By differentiating Eq.(\ref{func_eq02}) with respect to $y$, we have
\begin{equation} \label{eq11}
xf'_{\alpha,\beta}(xy) = x^{\alpha}f'_{\alpha,\beta}(y)+\beta y^{\beta-1}f_{\alpha,\beta}(x)
\end{equation}
Putting $y=1$ in Eq.(\ref{eq11}), we have the following differential equation:
\begin{equation} \label{eq12}
xf'_{\alpha,\beta}(x) - \beta f_{\alpha,\beta}(x)   =- c_{\alpha,\beta} x^{\alpha},
\end{equation}
where we put $c_{\alpha,\beta} \equiv-f'_{\alpha,\beta}(1)$. Eq.(\ref{eq12}) can be deformed as follows.
$$
x^{\beta+1}  \left(x^{-\beta}f_{\alpha,\beta}(x) \right)'  =- c_{\alpha,\beta} x^{\alpha}.
$$
That is, we have,
$$
 \left(x^{-\beta}f_{\alpha,\beta}(x) \right)' = -c_{\alpha,\beta} x^{\alpha-\beta-1}.
$$
Integrating both sides on the above equation with respect to $x$, we have
$$
 x^{-\beta}  f_{\alpha,\beta}(x) =
 -\frac{c_{\alpha,\beta}}{\alpha-\beta} x^{\alpha-\beta} +d_{\alpha,\beta},
$$
where $d_{\alpha,\beta}$ is a integral constant depending on $\alpha$ and $\beta$.
Therefore we have
$$
  f_{\alpha,\beta}(x) =  -\frac{c_{\alpha,\beta}}{\alpha-\beta} x^{\alpha} +d_{\alpha,\beta}x^{\beta},
$$ 
By $f_{\alpha,\beta}(1)=0$, we have $d_{\alpha,\beta}=\frac{c_{\alpha,\beta}}{\alpha-\beta}$.
Thus we have 
$$
 f_{\alpha,\beta}(x) = \frac{c_{\alpha,\beta}}{\alpha-\beta}\left(x^{\beta}-x^{\alpha}\right).
$$
Also by $f_{\alpha,\beta}(x) \geq 0$, we have $c_{\alpha,\beta} \geq 0$.

As for the case of $\alpha=\beta$, we can prove by the similar way.

\hfill \qed

If we take $\alpha=q,\beta=1$ or $\alpha=1,\beta=q$ in Theorem \ref{the02}, we have the following corollary.

\begin{Cor} \label{cor02}
If the differentiable nonnegative function $f_q$ with a positive parameter $q \in \mathbb{R}$ satisfies the following functional equation:
\begin{equation} \label{func_eq03}
f_q(xy)  = x^qf_q(y)+  y f_q(x) , \qquad (0< x,y \leq 1,\quad q\neq 1)
\end{equation}
then the function $f_q$ is uniquely given by
$$
f_q(x) = -c_q  x^q \ln_q x
$$
where $c_q$ is a nonnegative constant depending only on the parameter $q$.
\end{Cor}

Here we give an interpretation of the functional equation (\ref{func_eq03}) from the view of Tsallis statistics.

\begin{Rem}
Replacing $x$ by $y$ each other in Eq.(\ref{func_eq03}) and summing two functional equations, 
we have
\begin{equation} \label{func_eq03-1}
f_q(xy)  = \left(\frac{x^q +x}{2}\right) f_q(y)+ \left(\frac{y^q+ y}{2}\right) f_q(x) , \qquad (0< x,y \leq 1,\quad q\neq 1).
\end{equation}
Then the functional equation (\ref{func_eq03-1}) can be regarded as the sum of two following functional equations:
\begin{eqnarray}
&& f_q(xy)=yf_q(x)+xf_q(y)+(1-q)f_q(x)f_q(y) \label{eq21}\\
&&  f_q(xy)=y^qf_q(x)+x^qf_q(y)+(q-1)f_q(x)f_q(y) \label{eq22}
\end{eqnarray}
Two equations (\ref{eq21}) and (\ref{eq22}) imply the following equations for $i=1,\cdots,n$ and $j=1,\cdots,m$.
\begin{eqnarray}
&& f_q(x_iy_j)=y_jf_q(x_i)+x_if_q(y_j)+(1-q)f_q(x_i)f_q(y_j) \label{eq23}\\
&&  f_q(x_iy_j)=y_j^qf_q(x_i)+x_i^qf_q(y_j)+(q-1)f_q(x_i)f_q(y_j) \label{eq24}
\end{eqnarray}
Taking the sum for Eq.(\ref{eq23}) and Eq.(\ref{eq24}) on $i$ and $j$ with simple calculations, we have two
non-additivity relations given in Eq.(\ref{non-additive01}) and  Eq.(\ref{non-additive02}),
where we put $c_q =1$ for a simplicity.
Therefore we can conclude that two functional equations (\ref{eq21}) and  (\ref{eq22}), which are the essential parts 
of the non-additivity relations Eq.(\ref{non-additive01}) and  Eq.(\ref{non-additive02}),
characterize the Tsallis entropy function. In other words, the Tsallis entropy function can be characterized by two
non-additivity relations Eq.(\ref{non-additive01}) and  Eq.(\ref{non-additive02}).
\end{Rem}

If we again take the limit as $q \to 1$ in Corollary \ref{cor02}, we have the following corollary.
\begin{Cor} \label{cor03}
If the differentiable nonnegative function $f$ satisfies the following functional equation:
\begin{equation} \label{func_eq04}
f(xy)  =yf(x)+  xf(y) , \qquad (0< x,y \leq 1)
\end{equation}
then the function $f$ is uniquely given by
$$
f(x) = -c  x \log x
$$
where $c$ is a nonnegative constant.
\end{Cor}

\section*{Ackowledgement}
The author was partially supported by the Japanese Ministry of Education, Science, Sports and Culture, 
Grant-in-Aid for Encouragement of Young Scientists (B) 20740067.

\end{document}